\documentclass[12pt,twoside]{article}
\usepackage{fleqn,espcrc1}
\usepackage{latexsym} 
\usepackage{amssymb}  
\usepackage{amsfonts} 
\usepackage{epsf}
\usepackage{graphicx}
\def\y0{y^{(0)}}

\newcommand \beq{\begin{eqnarray}}
\newcommand \eeq{\end{eqnarray}}

\newcommand{\mnote}[1]{\marginpar{\tiny {}}}   

\begin{document}

\title{\bf (Non)Thermal Aspects of Charmonium Production and a New
Look at J/$\psi$ Suppression 
}
\author{P. Braun-Munzinger\address{Gesellschaft f\"ur
Schwerionenforschung, D 64291 Darmstadt, 
Germany}, J. Stachel\address{
Physikalisches Institut der Universit\"at Heidelberg, D 69120
Heidelberg, Germany }
\thanks{ The authors are grateful to the US Department of Energy's
Institute for Nuclear Theory at the University
of Washington for generous hospitality and support during a stay there
where part of this work was done.}
}


\maketitle

\begin{abstract}

    \noindent To investigate a recent proposal that J/$\psi$
    production in ultra-relativistic nuclear collisions is of thermal
    origin we have reanalyzed the data from the NA38/50 collaboration
    within a thermal model including charm. Comparison of the
    calculated with measured yields demonstrates the non-thermal
    origin of hidden charm production at SPS energy. However, the
    ratio $\psi^{'}$/(J/$\psi)$ exhibits, in central nucleus-nucleus
    collisions,  thermal features which lead us to a new interpretation of open
    charm and charmonium production at SPS energy. Implications 
    for RHIC and LHC energy measurements will be discussed.

\end{abstract}


\vspace{1.5cm}

The suppression of J/$\psi$ mesons (compared to what is expected from
hard scattering models) was early on predicted \cite{matsui} to be a
signature for color deconfinement. Data for S-induced collisions
exhibited a significant suppression but systematic studies soon
revealed that such suppression exists already in p-nucleus collisions
and is due to the absorption in (normal) nuclear matter of a
pre-resonant  state consisting, e.g., of a color
singlet $c \bar c g$ state that is formed on the way towards J/$\psi$
production. The situation has been summarized in
\cite{lourenco,kharzeev,satz}. 

The newest data for Pb+Pb collisions now exhibit clear evidence for anomalous
absorption beyond the standard nuclear absorption expected for such
systems. The most recent results are summarized in
\cite{na50_1,na50_2,cicalo}. The observed anomalous suppression is not
explained in  conventional models where the charmonia are broken up by
interactions with co-movers as discussed in \cite{na50_1}. For a
discussion of the present status of  J/$\psi$ suppression and its
understanding in terms of phenomenological models see \cite{blaizot}. 

However, it was recently conjectured \cite{gaz1,gaz2} that J/$\psi$
production is of thermal origin and exhibits no direct connection to
color deconfinement. Since the charmonia are heavy mesons with masses
much larger than any conceivable temperature, thermal production would
be a big surprize. On the other hand, substantial evidence now exists
\cite{therm3,therm2,therm1,stock,heinz,becattini,cleymans} that hadron
production (other than charmonia) in ultra-relativistic nuclear
collisions proceeds through a state of chemical equilibrium near or at
the phase boundary between hadron matter and quark-gluon plasma. For a
review of the implications for quark-matter physics see
\cite{pbmqm97,jsinpc,pbmpanic}.

To shed more light on the situation we have modified the thermal model
used in \cite{therm3} to include charmed hadrons. We will first
present an overview of the relevant modifications along with a
comparison of resulting thermal yields for J/$\psi$ mesons and of the 
$\psi^{'}$/(J/$\psi)$ ratio with experimental results from the NA38/50
collaboration. These results and a comparison to charm production in
hard scattering models suggest a new approach towards understanding
charmonium production in ultra-relativistic nuclear collisions, which
is discussed in the following. The implications for collider experiments
at RHIC and LHC will also be summarized.

The present statistical model \cite{therm3}  is based on the use of
a grand canonical ensemble 
to describe the partition function and hence the density of the
particles of  species \(i\) in an equilibrated fireball:

\begin{equation}
n_i= \frac{g_i}{2 \pi^2} \int_0^\infty \frac{p^2 \, {\rm
d}p}{e^{(E_i(p)-\mu_i)/T} \pm 1}
\label{grundgl}
\end{equation}

\noindent with \(n_i\) = particle density, \(g_i\) = spin degeneracy,
\(\hbar\) = c = 1, \(p\) = momentum, and \(E\) = total energy. The chemical
potential including charm degrees of freedom is written as  \(\mu_i =
\mu_B B_i+\mu_S S_i+\mu_{I_3} I^3_i +\mu_C C_i\). The 
quantities \( B_i\), \( S_i\), \( I^3_i\), and \(C_i\) are the baryon,
strangeness, three-component of isospin, and charm  quantum
numbers of the particle of 
species \(i\). The temperature T and the baryochemical potential
\(\mu_B\) are the two independent parameters of the model, while  the
strangeness chemical potential \(\mu_S\), 
the charm chemical potential \(\mu_C\), and the isospin chemical
potential \(\mu_{I_3}\) are fixed by strangeness, charm,
and charge conservation.  In addition, the volume V of the fireball is
determined by baryon conservation via the relation $n_{baryon} \rm{V}
=$ N$_{part}$, where N$_{part}$ denotes the number of  nucleons
participating in the collision and $n_{baryon}$ is the net baryon
density computed in the thermal model. 

In addition to the standard hadronic mass spectrum of 191 hadrons as used in
\cite{therm3} we have added mesons and baryons with open and hidden
charm. Specifically, open charm particles included are: D$^+$, D$^-$,
D$^0$, $\bar{{\rm D}^0}$, $\Lambda_c$, $\Sigma_c$, $\Lambda_c^*$,
$\bar{\Lambda_c}$, $\bar{\Sigma_c}$, $\bar{\Lambda_c^*}$, along with
the charmonia $\eta_c$, J/$\psi$, $\chi_0$, $\chi_1$, $\chi_2$,
$\psi^{,}$, $\psi^{,,}$, $\psi^{,,,}$.

\begin{figure}[thb]

\vspace{-1cm}

\epsfxsize=18cm
\begin{center}
\hspace*{0in}
\epsffile{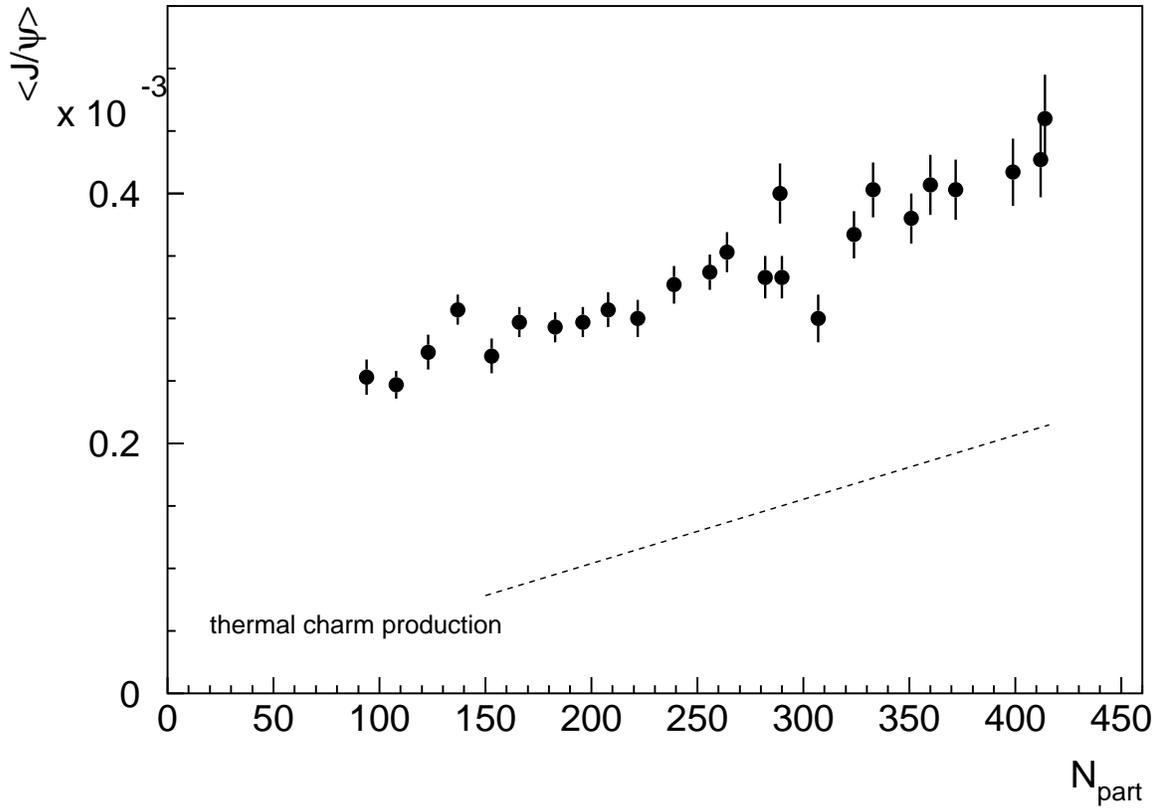}
\end{center}

\vspace{-2.5cm}

\caption{
Comparison of measured J/$\psi$ yield per Pb+Pb collision as a
function of centrality with the predictions of the
thermal model (dashed line). The data are from
\cite{bellaiche,b_et95} as analyzed in \cite{gosset1}.  For details see text}
\label{fig:jpsi_therm}
\end{figure}

Since the inclusion of these hadrons will modify the rest of the
hadron yields only at the sub-percent level, we will use, for the
following investigations, the temperature T= 168 MeV and baryon
chemical potential $\mu_B = 266$ MeV as established for central Pb+Pb
collisions at SPS energy in \cite{therm3}. This leads to $\mu_S$ = 71
MeV, $\mu_{I_{3}}$ = - 5 MeV, (both as in \cite{therm3}), and $\mu_C$
= - 65 MeV. Using a volume of 3085 fm$^3$ determined by baryon
conservation for central Pb+Pb collisions corresponding to 400
participants, we have compared, in Fig.~\ref{fig:jpsi_therm}, the
predictions of this thermal model with the data from NA50
\cite{bellaiche,b_et95}, as recently analyzed by J. Gosset et
al. \cite{gosset1} . Here the data for J/$\psi$ multiplicities are
plotted {\it vs} N$_{part}$. For this conversion we used the relation between
transverse energy and impact parameter as given for the 1995 NA50 data
in \cite{gosset1,b_et95} and the connection between impact parameter
and N$_{part}$ as established in \cite{b_npart}.

The predictions of the thermal model are represented by the dashed
line, where we have made use of the fact that, within the range of
applicability of the thermal model, all yields scale proportional to
the volume, i.e. proportional to the number of participants. We note
that, even for the most central collisions (where J/$\psi$ production
should be most suppressed \cite{matsui,blaizot}), the measured
J/$\psi$ yield is 
underpredicted by the thermal model calculations by more than a factor
of 2 and the discrepancy is about a factor of 3 for N$_{part}$ =
200. To compensate this factor of 3 discrepancy by an increase in
temperature\footnote{From T = 168 MeV to T=178 MeV, the J/$\psi$ yield
increases by a factor 2.7, the (J/$\psi)$/$\pi^-$ yield increases by a
factor of 2, and the $\psi^{'}$/(J/$\psi$) ratio by a factor 1.2.}
would require a temperature of T = 180 MeV, a value not compatible with that
determined from hadron production yields \cite{therm3}, where a
temperature range of 168 $\pm 4$ MeV (including systematic
uncertainties) was established.

We further note that it is highly doubtful that full chemical
equilibration can be reached for charmed hadrons at SPS
energies, either in a hadronic or a quark-gluon plasma scenario. Cross
sections for open charm production among hadrons are 
in the sub-$\mu \rm{b}$ level for relevant (thermal) energies, as is
estimated from \cite{charm_sps}. At such cross section levels the
equilibration times for charm should exceed those for strangeness,
where production cross sections exceed 100 $\mu \rm{b}$, by
more than 2 orders of magnitude. Taking into account that strangeness
equilibration times in a hadronic fireball exceed 50 fm/c \cite{life},
chemical equilibrium for charm in the hadronic sector can be ruled
out. Even in a quark-gluon 
plasma, where cross sections for charm production are much larger,
thermal production is small. Assuming a charm quark mass of 1.5 GeV
and an initial temperature of 300 MeV, very high for SPS energies,
Redlich has estimated  in a parton cascade approach \cite{redlich1,redlich2}
that, at hadronization (with T$_c$ = 160 MeV), the number of thermal c$\bar 
{\rm c}$ pairs is less than 0.01 in central Pb+Pb collisions, lower by
a factor of 40 of what would be needed to explain the
data. Furthermore, at T = 160 MeV, before the start of the mixed
phase, we estimate the volume of the plasma phase to be V$_{plasma} =
\pi R^2 \tau$ = 950 fm$^3$, assuming as in \cite{redlich1} a lifetime
of the plasma phase of 6.7 fm.  The number of charm quark pairs in chemical
equilibrium is then N$_{\rm{c} \bar{\rm{c}}}^{eq} = 0.47$, implying
that in the cascade approach the parton 
gas is undersaturated in charm by about a factor of 50!

In Table~\ref{tab:therm_yields} we present a summary of the results
from the thermal model calculation for mesons and baryons with open
and hidden charm\footnote{The thermal fluctuations about these mean
values n$^{therm}$ are Poisson distributed. This implies that per
collision the variance equals n$^{therm}$. For 10$^6$ Pb+Pb collisions, the
thermal prediction is, consequently, that 200 $\pm 14$ J/$\psi$ mesons will
be produced.}. It is interesting to compare these numbers with
predictions for the production of hadrons with open charm  in hard collisions.
Results obtained by PYTHIA calculations following \cite{charm_sps} are shown in
Table~\ref{tab:pythia_yields}. We  note that, somewhat
surprizingly, the yield of directly (via hard collisions) produced
charm is slightly larger than that produced if charm is in full
chemical equilibrium at T= 168 MeV. This implies that one cannot, under
any circumstances, neglect direct production. On the other hand, ratios
of charmed meson yields differ significantly from the thermal to the
direct scenario, as is obvious from  Table~\ref{tab:therm_yields} and
Table~\ref{tab:pythia_yields}.

Another interesting point concerns the $\psi^{'}$/(J/$\psi)$ ratio. As is
well known \cite{na50_3}, this ratio is, in hadron-proton and
p-nucleus collisions, close to 12 \%, independent of collision system,
energy, transverse momentum etc. In the thermal model, the ratio is
3.7 \%, including feeding of the J/$\psi$ from heavier charmonium
states. A temperature of about 
280 MeV would be necessary to explain the ratio found in pp  and
p-nucleus collisions
in a thermal approach. Clearly, J/$\psi$ and $\psi^{'}$ production in pp and
p-nucleus collisions  are manifestly non-thermal. This was previously
realized by Gerschel \cite{gerschel}. Similar
considerations apply for the $\chi$ states. In fact, feeding from
$\chi_1$ to J/$\psi$ is less than 3 \% if the production ratios are thermal.

\begin{figure}[thb]

\vspace{-1cm}

\epsfxsize=18cm
\begin{center}
\hspace*{0in}
\epsffile{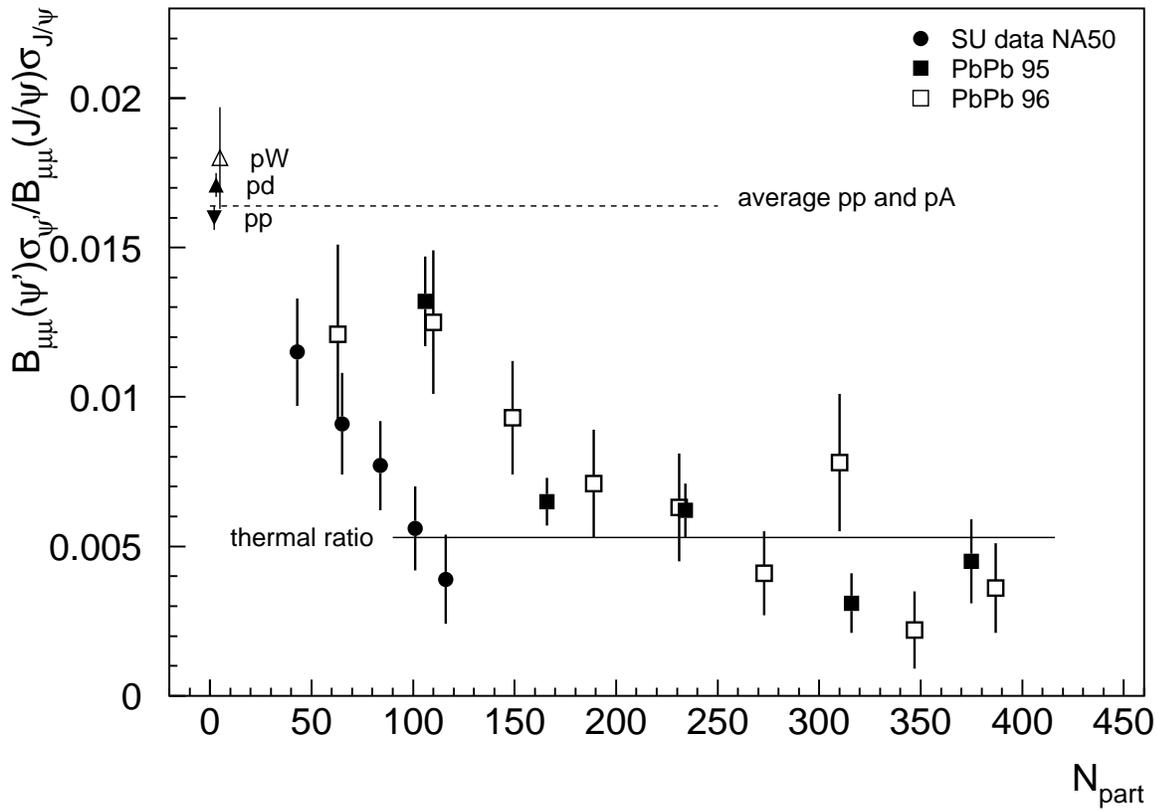}
\end{center}

\vspace{-1.5cm}

\caption{ Comparison of the dependence of the measured
$\psi^{'}$/(J/$\psi)$ ratio 
on the number of participating nucleons with the prediction of the
thermal model. The data are from the NA38/50 collaboration
\cite{na38_1,na50_4,na50_5,na38_2}. See text for more details.
}
\label{fig:psi'}
\end{figure}

The evolution with participant number of the $\psi^{'}$/(J/$\psi)$
ratio in nucleus-nucleus collisions is presented in
Fig.~\ref{fig:psi'}. The data are from the NA38/50 collaboration
\cite{na38_1,na50_4,na50_5,na38_2}. With increasing N$_{part}$ the
$\psi^{'}$/(J/$\psi)$ ratio drops first rapidly (away from the value
in pp collisions) but seems to saturate for high N$_{part}$ values at
a level very close to the thermal model prediction, both for S+U and
Pb+Pb collisions

This surprizing fact along with the previous observations leads us to
propose a new scenario for J/$\psi$ production. We assume that {\bf
all} c$\bar {\rm c}$ pairs are produced in direct, hard collisions,
i.e. in line with previous considerations we neglect thermal
production. For a description of the hadronization of the c and $\bar
{\rm c}$ quarks, i.e. for the determination of the relative yields of
charmonia, and charmed mesons and baryons, we employ the statistical
model, with parameters as determined by the analysis of all other
hadron yields \cite{therm3}. The picture we have in mind is that all
hadrons form within a narrow time range at or close to the phase
boundary. 

Since the number of directly produced charm quarks deviates from the
value determined by chemical equilibration, we introduce a charm
enhancement factor $g_c$ by the requirement of charm
conservation. This leads to:

\beq
{N_{c\bar c}^{direct} = \frac{1}{2} g_cV(\sum_{i}
n_{D_i}^{therm}+n_{\Lambda_i}^{therm})+
g_c^2V(\sum_{i}n_{ \psi_i}^{therm}) +...}.
\eeq

Via this equation the thermal yields (thermal densities times volume
V) are adjusted to the yield of directly produced charm quark
pairs. The remaining terms in eq. (2) of second order and third  order in $g_c$
are completely negligible. Of course
this equation makes 
only sense as long as N$_{c\bar c}^{direct}$ is much less than the
number of up, down, and strange quarks in the fireball, a relation
which is well fulfilled up to the highest (LHC) energies considered
here. 

The number of D  and J/$\psi$ mesons are then enhanced relative to the
thermal model prediction by factors g$_c$ and g$_c^2$, i.e. 

\beq
N_D=g_cV n_{D}^{therm} \hspace{0.6cm} \rm{and} \hspace{0.7cm}
N_{J/\psi}=g_c^2V n_{J/\psi}^{therm}.
\eeq

\begin{figure}[thb]

\vspace{-1cm}

\epsfxsize=19cm
\begin{center}
\hspace*{-1.0cm}
\epsffile{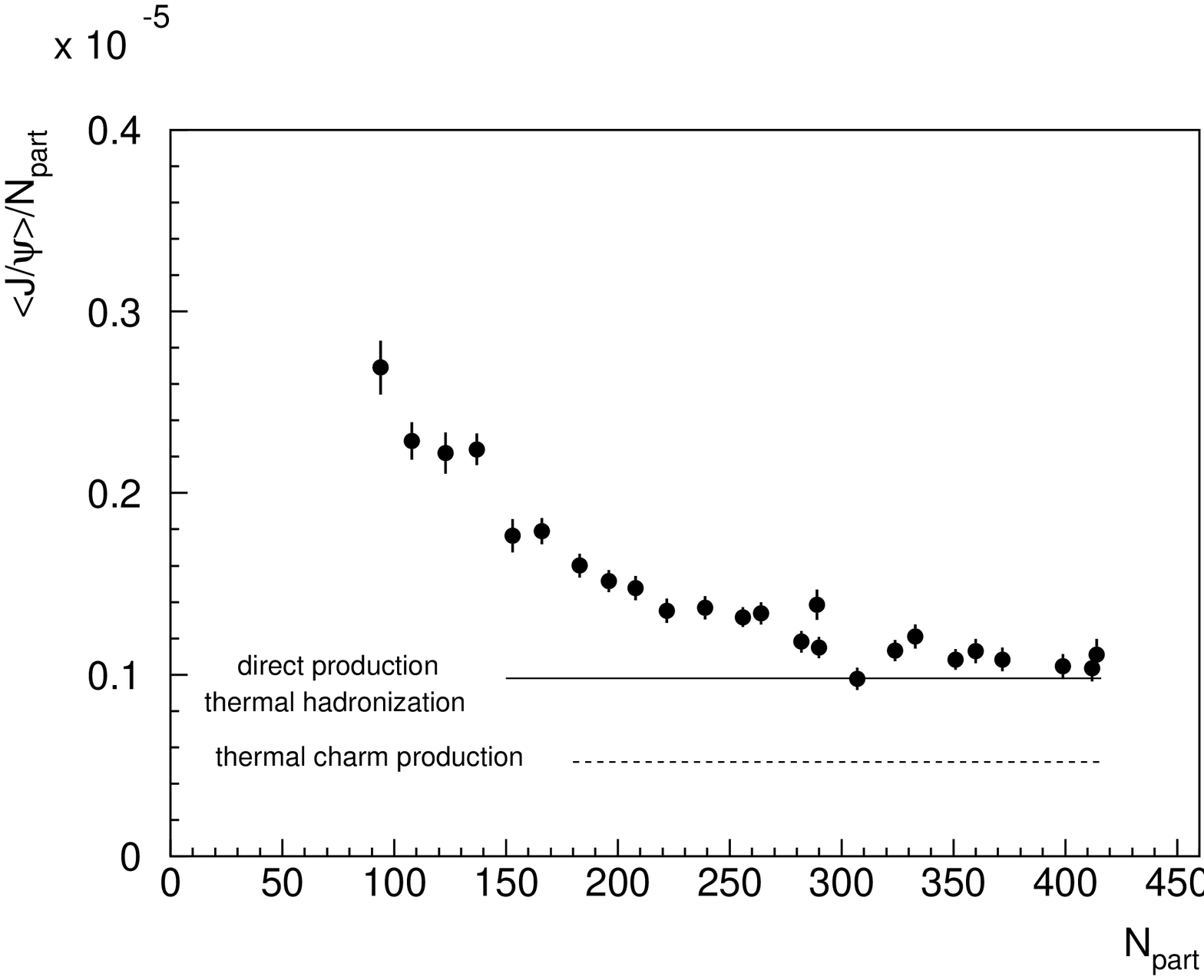}
\end{center}

\vspace{-1.5cm}

\caption{ Comparison of the dependence of the measured
(J/$\psi)$/N$_{part}$ ratio 
on the number of participating nucleons with the predictions of the
thermal model (dashed line) and of the direct/statistical model (solid
line) The data are from
\cite{gosset1,bellaiche,b_et95}. For details see text.
}
\label{fig:psipp}
\end{figure}

Using this approach of direct production and statistical hadronization
we have recalculated the yield for J/$\psi$ mesons. For N$_{part} =
400$ in Pb+Pb collisions the directly produced number of charm quark
pairs is $N_{c\bar c}^{direct}= 0.173$ per collision, using the PYTHIA
parameters of \cite{charm_sps}. This implies
g$_c = 1.38$. The resulting J/$\psi$ yield per participant is plotted,
in Fig.~\ref{fig:psipp} as a function of N$_{part}$ \footnote{For the
N$_{part}$ dependence of open charm we use the form deduced from the
dependence on transverse energy of the Drell-Yan yield, as determined
by \cite{gosset1}. Somewhat surprizingly, this scales approximately
linearly in N$_{part}$, not like N$_{part}^{4/3}$ as expected for hard
scattering.}.

Without introduction of new parameters this approach describes the
measured yield for J/$\psi$ mesons very well for those collision
centralities where also the $\psi^{'}$/(J/$\psi)$ ratio is well
described (see Fig.~\ref{fig:psi'}). The interpretation of these
results is as follows: at SPS energies charm is produced directly in
nuclear collisions at the rate expected by straight extrapolation from
what is known about charm in pp collisions \cite{charm_sps}. From
these charm quarks all charmed hadrons are formed by statistical hadronization.

We would like to point out, however, that the present approach is
rather schematic. First, the absolute yield of charm in
nucleus-nucleus collisions is not well known. From \cite{charm_sps} we
deduce an uncertainty of the K factor of about 2. Furthermore, 
open charm production might be  enhanced by a factor of 3 in Pb+Pb
collisions over expectations from pp collisions, as has been conjectured in
\cite{b_npart}. Finally, inclusion of additional charmed mesons (in
particular D$^*$'s), will modify the absolute yield of thermal charm. A
quantitative description will require a measurement of the open charm
yield in Pb+Pb collisions.

An interesting question was raised after the first submission of this
paper, namely whether the canonical or grand canonical partition
function should be used \cite{redlich_3}. In the present approach, we
have chosen the simpler grand-canonical ensemble. We remark that the
N$_{part}$ dependence of the J/$\psi$ yield may help to decide among
the different approaches.

In the present approach all J/$\psi$
mesons result, for the most central collisions, from the statistical
hadronization of the directly produced charm quarks. Directly produced
J/$\psi$ mesons are (i) not formed before the reaction proceeds into a
plasma phase or (ii) effectively destroyed during the plasma phase by,
e.g., a color screening mechanism as proposed in \cite{matsui}. In
either case the current interpretation requires the existence of a
deconfined phase during the collision. We remark here that Kabana
\cite{kabana} has recently argued that coalescence of charm quarks is
the source of J/$\psi$ mesons in nuclear collisions. While in spirit
this is similar to our approach, this scenario assumes an enhancement of
open charm which increases with N$_{part}$, very different from the
present conclusions. In the context of the above arguments it would be
very important to get a direct measurement of open charm production in
nucleus-nucleus collisions.

If the present scenario is correct, the consequences for quarkonium
production at collider energies could be significant. For RHIC
energies near central rapidity, e.g., the number of directly produced
charm quark pairs per unit rapidity is \cite{vogt_1} dN$_{c \bar
c}^{direct}$/dy =1.0, leading to g$_c$ = 6.8 and
dN$^{therm}_{J/\psi}$/dy = 10$^{-2}$, close to the unsuppressed value
expected from hard collisions \cite{vogt_2}. If the directly produced
charm quarks hadronize statistically, as is implied at SPS energy, we
would predict no J/$\psi$ suppression at all at RHIC energies, even
though there are no J/$\psi$ mesons during the plasma phase. At LHC
energy, the current approach would actually predict a significant
enhancement of charmonia over the value expected for direct
production. Of course, the underlying assumption is that the momenta
of the charm quarks are close to thermal near the critical
temperature, i.e. thermal (but not chemical) equilibration is
required. Present SPS data are not at variance with such a scenario,
since the measured transverse momentum spectra for J/$\psi$ 
mesons \cite{na38_pt,na50_5} exhibit thermal shapes with inverse slope
constants around 230 MeV, as expected for a heavy particle which
participates little in the transverse flow build-up during hadronic
expansion. Whether this thermalization will also take place at
collider energies is an interesting open question.

We finally note the difference of the present approach with that
described in \cite{redlich2}, where secondary charmonium production
during the mixed phase is calculated under the assumption that all
charm quark pairs end up in D mesons after hadronization. Since, even
in the present approach, the yield of charmonia is small compared to
the yield of D mesons, the calculations reported in \cite{redlich2}
concerning secondary charmonium production should still be valid. In
fact, the charmonium production yields considered there should be added to
the present predictions.

In summary, we have shown that assuming chemical  equilibration of charm
does not lead to a successful thermal description of available data for
p-nucleus and nucleus-nucleus collisions. However, the experimental
data at SPS energy for the ratios $\psi^{'}$/(J/$\psi)$ and
(J/$\psi)$/N$_{part}$ exhibit thermal features for the most central
Pb+Pb collisions. Coupled with the fact that direct production of
charm quarks  in hard scattering is close to the value obtained by
assuming full chemical equilibrium these observations lead to a new
interpretation of charmonium production in nucleus-nucleus collisions
in terms of a direct production and statistical hadronization
approach. This describes the SPS data well and suggests a possible
revision of the scenario for charmonium production at collider energies.

\newpage

\begin{table}[htb]
\newlength{\digitwidth} \settowidth{\digitwidth}{\rm 0}
\catcode`?=\active \def?{\kern\digitwidth}


\caption{Yields for particles with open and hidden charm
at mid-rapidity for central Pb+Pb 
collisions as predicted by the thermal model calculation.
Chemical freeze-out takes place at T=168 MeV in a volume of 3085
fm$^3$ corresponding to 400 participants.
The chemical potentials are: $\mu_B = 266$ MeV, $\mu_S = 71$ MeV, $\mu_C
=-65$ MeV, $\mu_{I_3} = - 5$ MeV. The effect of feeding of the
J/$\psi$ meson from heavier charmonia, including the relevant decay
branching ratios, is shown in the last row.}

\vspace{1.0cm}

\label{tab:therm_yields}
\begin{center}
\begin{tabular}{||c|c|c||}
\hline
& & \\
particle species & yield/$\pi^-$ & total \\
&  & \\
\hline
\(\rm{D}^+ =\rm{D}^0  \)            & 5.8$\cdot 10^{-5}$    & 0.034 \\
\(\rm{D}^- =\overline{ \rm{D}^0} \) & 1.3$\cdot 10^{-4}$    & 0.073 \\
\(\rm \Lambda_c \)                  &  6.4$\cdot 10^{-5}$     & 0.036 \\
\(\rm \overline{\Lambda_c} \)       &  5.8$\cdot 10^{-6}$     & 3.3
$\cdot 10^{-3}$ \\
\hline
\(\rm{J}/\psi \)                    & 3.45 $10^{-7}$         &2.0 $\cdot
10^{-4}$ \\
\(\chi_1\)                          & 3.5 $10^{-8}$         & 2.0$\cdot
10^{-5}$ \\          
\(\psi'\)                            & 1.3 $10^{-8}$         &7.6
$\cdot 10^{-6}$ \\
\(\rm{J}/\psi +\chi's + \psi'\)     & 3.6 $10^{-7}$         &2.1 $\cdot
10^{-4}$ \\
\hline
\hline
\end{tabular}
\end{center}

\end{table}

\begin{table}[htb]
\catcode`?=\active \def?{\kern\digitwidth}

\vspace{-0.0cm}

\caption{Yields for particles with open charm for central Pb+Pb 
 collisions as predicted by PYTHIA calculations \cite{charm_sps}  for
 N-N collisions and 
 scaled to central (N$_{part}$ = 400) Pb+Pb collisions using the nuclear
 thickness function.  } 

\vspace{1.0cm}

\label{tab:pythia_yields}
\begin{center}
\begin{tabular}{||c|c||}
\hline
&  \\
particle species  & total yield \\
&  \\
\hline
\(\rm{D}^+  \)               & 0.032 \\
\(\rm{D}^- \)     & 0.036 \\
\(\rm{D}^0 \)  &  0.093 \\
\(\overline{ \rm{D}^0} \) & 0.12 \\
\(\rm \Lambda_c \)                     & 0.034 \\
\(\rm{D}^-_s \)     & 0.011 \\
\(\rm{D}^+_s \)     & 0.015 \\
\hline
\hline
\end{tabular}
\end{center}

\end{table}

\newpage

\begin{thebibliography}{9}
\vspace{3mm}



\bibitem{matsui} T. Matsui and H. Satz, Phys. Lett. {\bf B178} (1986) 416.


\bibitem{lourenco} C. Louren\c{c}o, Nucl. Phys. {\bf A610} (1996) 552c.

\bibitem{kharzeev} D. Kharzeev, Nucl. Phys. {\bf A638} (1998) 279c.

\bibitem{satz} H. Satz, hep-ph/0007069.

\bibitem{na50_1} M.C. Abreu et al., NA50 collaboration, Phys. Lett. {\bf
B477} (2000) 28.


\bibitem{na50_2} M.C. Abreu et al., NA50 collaboration, Phys. Lett. {\bf
B450} (1999) 456.


\bibitem{cicalo} C. Cicalo for the NA50 collaboration, Proc. Quark
Matter 99 conference, Torino, June 1999, Nucl. Phys. {\bf A661} (1999) 93c.

\bibitem{blaizot} J.P. Blaizot, M. Dinh, J.Y. Ollitrault, nucl-th/0007020.

\bibitem{gaz1} M. Gazdzicki, Phys. Rev. {\bf C60} (1999) 054903.
\bibitem{gaz2} M. Gazdzicki and M. Gorenstein, Phys. Rev. Lett. {\bf
83} (1999) 4009.

\bibitem{therm3} P. Braun-Munzinger, I. Heppe, J. Stachel,
Phys. Lett {\bf B465} (1999) 15. 


\bibitem{therm2} P. Braun-Munzinger, J. Stachel, J. P. Wessels,
N. Xu, Phys. Lett.  {\bf B365} (1996) 1.  


\bibitem{therm1}P. Braun-Munzinger, J. Stachel, J. P. Wessels, N. Xu,
Phys. Lett.  {\bf B344} (1995) 43.  



\bibitem{stock} R. Stock, Phys. Lett. {\bf B456} (1999) 277.

\bibitem{heinz} U. Heinz, Nucl. Phys. {\bf A661} (1999) 140c.

\bibitem{becattini} F. Becattini, M. Gazdzicki, J. Sollfrank,
Eur. Phys. J. {\bf C5} (1998) 143.

\bibitem{cleymans} J. Cleymans, K. Redlich, Phys. Rev. {\bf C60} (1999) 054908.

\bibitem{pbmqm97} P. Braun-Munzinger and J. Stachel, Nucl. Phys. {\bf A638}
(1998) 3c.


\bibitem{jsinpc} J. Stachel, Proc. INPC, Paris, August 1998,
Nucl. Phys. {\bf A654} (1999) 119c.

\bibitem{pbmpanic} P. Braun-Munzinger, Proc. PANIC, Uppsala, June
1999, Nucl. Phys. {\bf A663-664} (2000) 183.


\bibitem{bellaiche} F. Bellaiche, dissertation, Univ. Claude Bernard,
Lyon 1, 1997. 

\bibitem{b_et95} M. C. Abreu et al., NA50 collaboration,
Phys. Lett. {\bf B410} (1997) 337.


\bibitem{gosset1} J. Gosset, A. Baldisseri, H. Borel, F. Staley,
Y. Terrien, Eur. Phys. J. {\bf C13} (2000) 63.


\bibitem{b_npart} M. C. Abreu et al., NA50 collaboration,
Eur. Phys. J. {\bf C14} (2000) 443.


\bibitem{charm_sps} P. Braun-Munzinger, D. Miskowiec, A. Drees,
C. Louren\c{c}o, Eur. Phys. J. {\bf C1} (1998) 123.

\bibitem{life} J. Sollfrank and U. Heinz in: Quark Gluon Plasma 2,
R.C. Hwa, editor, World Scientific 1996, p. 555.

\bibitem{redlich1} K. Redlich, private communication.

\bibitem{redlich2} P. Braun-Munzinger and K. Redlich, hep-ph/0001008
and Eur. Phys. J  {\bf C} (in print).

\bibitem{na50_3} M.C. Abreu et al., NA50 collaboration, Phys. Lett. {\bf
B438} (1998) 35, and references therein.

\bibitem{gerschel} C. Gerschel, Acta Phys. Pol. {\bf B30} (1999) 3585.

\bibitem{na38_1} M.C. Abreu et al., NA38 collaboration, Phys. Lett. {\bf
B449} (1999) 128.

\bibitem{na50_4} M. Gonin et al., NA50 collaboration, Proc. 3rd
Conf. on Physics and Astrophysics of Quark-Gluon Plasma, Jaipur,
India, March 1999, B. C. Sinha, D.K Srivastava, Y.P. Viyogi, editors,
Narosa Publ. House 1998, p. 393.


\bibitem{na50_5} L. Ramello et al., NA50 collaboration,
Nucl. Phys. {\bf A638} (1998) 261c.

\bibitem{na38_2}  M.C. Abreu et al., NA38 collaboration, Phys. Lett. {\bf
B466} (1999) 408.

\bibitem{redlich_3} K. Redlich and B. M\"uller, private communication.


\bibitem{kabana} S. Kabana, hep-ph/0004138.

\bibitem{vogt_1} V. Emel'yanov, A. Khodinov, S.R. Klein, R. Vogt,
Phys. Rev. Lett. {\bf 81} (1998) 1801.

\bibitem{vogt_2} V. Emel'yanov, A. Khodinov, S.R. Klein, R. Vogt,
Phys. Rev. {\bf C61} (2000) 044904.

\bibitem{na38_pt} M.C. Abreu et al., NA38 collaboration, Phys. Lett. {\bf
B423} (1998) 207.


\end{thebibliography}
\end{document}